\documentclass[conference]{IEEEtran}
\IEEEoverridecommandlockouts
\usepackage{multirow}
\usepackage{cite}
\usepackage{amsmath,amssymb,amsfonts}
\usepackage{algorithmic}
\usepackage{graphicx}
\usepackage{textcomp}
\usepackage{xcolor}
\usepackage{subfigure} 
\usepackage{float}
\usepackage{booktabs}
\usepackage[colorinlistoftodos]{todonotes}
\usepackage{tabularray}
\usepackage{array}
\usepackage{pdflscape}
\usepackage{longtable}
\usepackage{geometry}
\def\BibTeX{{\rm B\kern-.05em{\sc i\kern-.025em b}\kern-.08em
    T\kern-.1667em\lower.7ex\hbox{E}\kern-.125emX}}

\begin{document}

\title{Beyond Rankings: Exploring the Impact of SERP Features on Organic Click-through Rates\thanks{\footnotesize \textsuperscript{*}Note: these authors contributed equally to this work and shared first authorship
}}

\author{
    \IEEEauthorblockN{Erik Fubel\IEEEauthorrefmark{1}, Niclas Michael Groll\IEEEauthorrefmark{1}, Patrick Gundlach\IEEEauthorrefmark{1}, Qiwei Han\IEEEauthorrefmark{1},
    Maximilian  Kaiser\IEEEauthorrefmark{2}}
    \IEEEauthorblockA{\IEEEauthorrefmark{1}Nova School of Business and Economics, Carcavelos, Portugal
    \\\{49155, 48699, 49594, qiwei.han\}@novasbe.pt}
    \IEEEauthorblockA{\IEEEauthorrefmark{2}Universität Hamburg, Hamburg, Germany 
    \\{maximilian.kaiser@uni-hamburg.de}}
}

\maketitle

\begin{abstract}
Search Engine Result Pages (SERPs) serve as the digital gateways to the vast expanse of the internet. Past decades have witnessed a surge in research primarily centered on the influence of website ranking on these pages, to determine the click-through rate (CTR). However, during this period, the landscape of SERPs has undergone a dramatic evolution: SERP features, encompassing elements such as knowledge panels, media galleries, FAQs, and more, have emerged as an increasingly prominent facet of these result pages. Our study examines the crucial role of these features, revealing them to be not merely aesthetic components, but strongly influence CTR and the associated behavior of internet users. We demonstrate how these features can significantly modulate web traffic, either amplifying or attenuating it. We dissect these intricate interaction effects leveraging a unique dataset of 67,000 keywords and their respective Google SERPs, spanning over 40 distinct US-based e-commerce domains, generating over 6 million clicks from 24 million views. This cross-website dataset, unprecedented in its scope, enables us to assess the impact of 24 different SERP features on organic CTR. Through an ablation study modeling CTR, we illustrate the incremental predictive power these features hold.
\end{abstract}


\section{Introduction}\label{introduction}
Search Engine Optimization (SEO) is a strategic process used in online marketing with the aim of enhancing a website's visibility in search engine results pages (SERPs). The objective is to attract the highest possible volume of organic traffic to a website, which is a cornerstone strategy for long-standing online marketing \cite{ledford2009search}. In the realm of e-commerce, it's estimated that 33\% of total web traffic originates from organic search results \cite{chevalier2022globalb}. Organic traffic encompasses those instances where a user inputs a keyword in a search engine and lands on a website by clicking on one of the non-ad results. Given the high relevance of organic traffic, it is crucial for website providers to optimize their strategies to maximize the likelihood of generating clicks through search engines. Generally, the clicks to a website on a search result page for a specific keyword can be formulated as follows:
\begin{equation} \label{equation:clicks}
    Clicks=Impressions * CTR
\end{equation}

where CTR refers to the click-through rate of a result \cite{Google2022a}. To enhance the CTR, SEO efforts usually center on a website's rank on a result page, as top-ranked websites command more user attention and consequently receive more clicks \cite{lewandowski2020influence}. To secure a high rank on a result page, a website needs to demonstrate relevance for the particular keyword. Achieving this typically involves aligning a website’s content and metadata with the keyword, a focus for many SEO practitioners \cite{cui2011search,shih2013retracted,bala2018,ziakis2019important,das2021search,olson2021business}.

\begin{figure}[htbp]
\centerline{\includegraphics[width=\columnwidth]{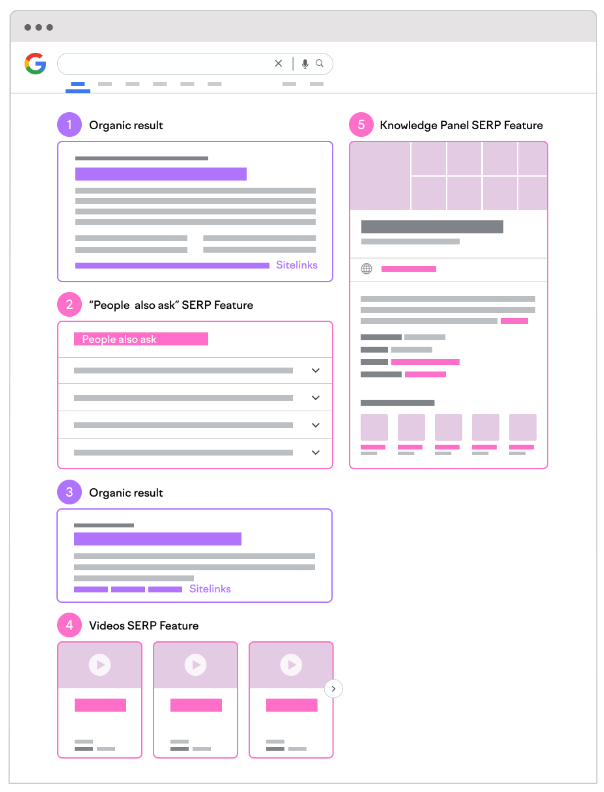}}
\caption{Exemplification of SERP features on a Google result page adapted from \cite{semrush2022serp}, between organic results (1) and (3), such as answer box (2), video results (4), knowledge panel (5), among others.}
\label{fig:serps_examples}
\end{figure}

Although introduced in the early 2000s, \textit{SERP features} — a visually prominent component of search results—have been heavily utilized by Google since the mid-2010s \cite{oliveira2023search_interface_evolution}. These SERP features are elements that encapsulate information from organic search results with the intent of making Google's result pages more engaging \cite{Google2022b}. They can take various forms, such as providing an instant answer to a user’s question, displaying a knowledge panel on the right of the result page, or including an image alongside a result \cite{oliveira2023serp_evolution}. Figure \ref{fig:serps_examples} provides an illustration of SERP features.

Introducing these visual elements can significantly influence users’ clicking decisions, thus impacting CTR and the revenue funneling into e-commerce websites. Some argue that by integrating additional features into its result pages and leveraging information originally published on third-party websites, Google reduces users' need to leave its ecosystem \cite{tober2022zero}. This could potentially disadvantage the publishers of the original information who rely on website traffic. In contrast, Google asserts that inclusion in SERP features can significantly boost CTR, visits, and time spent on a website \cite{Google2022b}. Despite the increasing prevalence of SERP features, their precise impact on CTR remains poorly understood. This gap in our knowledge raises an important research question: \textbf{To what extent do SERP features influence CTR, and how does their presence affect the importance of a website's ranking position?}

In this study, we broaden the existing analysis of the influence of a website's ranking position, which has been identified as the main direct influence on CTR, to include the characteristics of SERP features. This inclusion of SERP feature characteristics presents a novel aspect of CTR research. Furthermore, we compare the relative importance of SERP features with other result page characteristics.

This paper is structured as follows: Section \ref{related_works} offers a comprehensive review of related works on the influences on CTR, including research on SERP features and CTR prediction models. Section \ref{data} and \ref{methodology} provide detailed descriptions of the dataset used in this study and the methodology applied, respectively. Section \ref{eda} presents an exploratory data analysis of the interactions between SERP features, CTR, and other features. In Section \ref{modeling}, we investigate the importance of SERP features by examining their predictive power. Finally, Section \ref{conclusion} provides a discussion of the findings and concludes the paper.

\section{Related Works} \label{related_works}

\subsection{Determinants of CTR} \label{related_works:determinants_ctr}

Search engine optimization (SEO) is one of the most widely adopted online advertising strategies in the present day, and it plays a crucial role in driving organic website traffic \cite{olson2021business,kingsnorth2022digital}. A significant body of research in the field of SEO marketing has identified the ranking position of a website on a search engine results page (SERP) as the primary determinant of click-through rates (CTR), making it a central focus of SEO efforts \cite{cui2011search,shih2013retracted,bala2018,ziakis2019important,das2021search,olson2021business}. SEO strategies aim to elevate the rank of a website by meticulously tailoring its content to align with specific keywords. This might involve adding new, high-quality blog posts to improve the relevance of the website to search engines \cite{das2021search}. Other factors such as keyword characteristics or result characteristics have been studied predominantly for their impact on the position of a website in search results, and less so for their direct influence on CTR \cite{sheffield2020seo_interviews}. For instance, much of the literature on keywords aims to identify strong keywords and trends and recommends optimal keyword characteristics, content adaptation, linkage optimization, or structure adjustment to achieve higher rankings on search engine result pages \cite{erdman2011, Malaga2008, Nagpal2021, yalcin2010}. While the position of a website has been widely recognized as the primary driver of CTR, it is equally important to consider additional characteristics such as result, keyword, and SERP features that directly influence CTR.

\subsection{SERP Features} \label{related_works:serp_features}

Although Google introduced the first SERP feature as early as 2002, it was not until the mid-2010s that their usage intensified \cite{oliveira2023serp_evolution}. Today, SERP features are among the most conspicuous and widely used elements on search engine result pages. Oliveira and Texeira Lopes provide a comprehensive overview of the evolution of SERP features on Google Search and Microsoft Bing, noting that these features have become more common and diverse, aggregating content from different verticals and providing more features that give direct answers \cite{oliveira2023serp_evolution}. However, critics argue that Google's addition of more SERP features and use of information originally published on third-party websites has reduced the need for users to leave Google's ecosystem, thereby disadvantaging the publishers of the original information who depend on traffic to their websites. According to a Semrush study, zero-click searches, where a user does not leave Google’s ecosystem, account for 25.6\% of all searches \cite{tober2022zero}.

Despite the long-standing existence of SERP features and their visual dominance on search engine result pages, their influence on user behavior has only been sporadically studied. For instance, one study analyzes the positive effect on CTR when a website appears in the featured snippet, but this analysis relies on information found on blogs from influential SEO companies \cite{sammartin2020google}. Numerous influential marketing blogs tout the benefits of appearing in SERP features, with many suggesting that SERP features can enhance the visibility of websites featured in them and reduce the CTR of websites that appear alongside SERP features without being featured \cite{moz2022serp,semrush2022serp,wheelhouse2022serp}. Google advertises significant improvements in click-through rate, visits, and time spent on websites when websites are chosen to be shown in SERP features  \cite{Google2022b}. Despite these claims, the actual effect of SERP features on CTR remains largely unexplored and under-theorized. This gap in the literature indicates that there is significant potential for this work to illuminate the opaque role of SERP features in Google’s search engine ecosystem and provide the first empirical research on the SERP features’ impact on website performance \cite{wheelhouse2022serp}.

\subsection{CTR Prediction Models} \label{related_works:models}
Generally, CTR prediction research has largely been concerned with binary classification problems, asking questions such as "Will user X click on item Y?" The items under consideration are typically ads on a search engine result page, but they can also be articles in an online shop \cite{yang2022click}. This approach differs fundamentally from the problem formulation in this work, which treats CTR prediction as a regression problem, predicting a continuous value based on aggregated data, as opposed to making a classification for a single observation. This difference is also reflected in the nature of the datasets and models used for classification. Datasets consisting of individual users and items are often highly sparse and high-dimensional after one-hot encoding \cite{yang2022click,zhou2018deep}. As a result, the research and the resulting models have been developed to handle this high dimensionality and sparsity \cite{zhou2018deep}.

The field of CTR prediction has seen the development of a range of models, including multivariate statistical models \cite{wang2013advertisement,yan2014coupled}, factorization machines and field-aware factorization machines \cite{ma2016f2m,pan2018field,yuchin2016field}, and combinations of deep learning models and factorization machines \cite{chen2016deep,cheng2016wide,guo2017deepfm,gharibshah2020deep,deng2021deeplight,wang2021masknet}. Yang provides a detailed overview of CTR prediction in the context of user-item interactions \cite{yang2022click}. However, these models, which were developed for classification on sparse ad-click datasets, are not suitable candidates for predicting aggregated organic CTR (see section \ref{methodology:examined_models}). While recent research has focused on predicting user-item interactions, to our knowledge, only one work has proposed a model for predicting aggregated CTR \cite{richardson2007predicting}. However, this work is concerned with the prediction of ad CTR rather than organic results. 

To summarize, although a substantial body of literature exists on CTR prediction, these works differ from the problem at hand in terms of the nature of prediction (individual vs. aggregated) and the type of clicks (organic results vs. ads). This suggests that the models proposed in these works may not be directly well-suited to the prediction problem in our study, especially in light of understanding the importance of SERP features.

\section{Dataset Description} \label{data}

\subsection{Data Acquisition and Preprocessing} \label{data:preprocessing}
The primary dataset employed for this study was provided by Grips, a German startup that endeavors to generate a comprehensive map of online commerce for retailers and brands \cite{grips2023about_us}. The dataset is an aggregation of historical Google search data, specifically gathered from US-based desktop searches during the period spanning May 31st, 2022, to August 18th, 2022. With more than 24 million views and over 6 million user clicks on search engine result pages, the dataset presents an extensive source of information. The data has been collected across approximately 67,000 distinct search terms, commonly referred to as keywords, and 43 different e-commerce stores representing a broad array of industries.

Each data point in the dataset provides insight about a specific URL and keyword, encapsulating metrics related to the performance of the URL for the corresponding keyword. To illustrate, the data could reflect metrics pertinent to the performance of the URL 'www.amazon.coom/' for the searched keyword 'amazon'. Thus, the data is presented in a tabular, heterogeneous format.

The primary dataset was further enhanced with additional features derived from a leading search engine marketing company and Google’s Keyword Planner. These features encompass a diverse range of aspects, from the searched keyword and metadata about the result page to information specific to the result and the displayed SERP features. The dataset also includes the click-through rate for each result. In its raw form, the dataset comprises 70 features, among which is information about the presence of 24 distinct SERP features (see the source of each feature in appendix \ref{appendix:data_dict}). The dataset also provides granular data about which specific result was included in which SERP feature. As such, given its extensive set of result page features, which encompass real-world, non-publicly accessible data for numerous e-commerce stores, this dataset represents a unique and innovative resource for research.

In order to prepare the data for subsequent analysis and modeling, the guidelines for data preprocessing as outlined by Google adhered to \cite{Google2022c,google2022d}. To reduce the potential for noise in the data due to the observations of result pages that received minimal user exposure, an impression threshold of 20 impressions per result was imposed. As a result, 58,898 data entries remain for analysis.

\subsection{Feature Categorization and Engineering} \label{data:feature_engineering_and_subsets}
To facilitate a comprehensive comparison of the impact of SERP features on CTR, in relation to other characteristics of the result page, the variables within the dataset were divided into three distinct categories or subsets: \textit{Position}, \textit{Keyword}, and \textit{SERP features}. Moreover, self-generated features were incorporated, where relevant, to augment these subsets. These engineered feature subsets are employed in both the overall data analysis and the modeling process. A brief overview of these subsets is provided below. For a comprehensive list of features within each subset, please see appendix \ref{appendix:subsets}, and for a detailed description of all features, please refer to the data dictionary in appendix \ref{appendix:data_dict}.

\begin{itemize}
\item \textbf{Position}: This subset includes features that are explicitly related to the ranked position of the result. These features encompass the current result position at the time of measurement, the monthly average position, and the positional difference from the previous measurement. While the position subset is treated as a distinct subset in the subsequent analysis, it is crucial to acknowledge that the position feature plays a pivotal role in addressing the problem at hand. As emphasized by the literature review and the findings of this study, the position is the most significant predictor of CTR. Therefore, the analysis will be conducted using an ablation study approach, using the position subset as a baseline and incorporating the additional subsets as components to assess their relative importance beyond that of position.

\item \textbf{Keywords}: This subset comprises the keyword itself and all directly resulting elements from the keyword a user inputs into the search engine. This includes details about the keyword, such as the length of the keyword, information indirectly related to that keyword like the level of competition or the search volume, and finally, the search intent. To provide a numerical representation of the complexity of each keyword, the Flesch reading ease score was computed \cite{flesch1948reading_ease}. Although more recent readability measures have been developed, Flesch’s reading ease score was chosen for its widespread acceptance and straightforward interpretability. This score is calculated based on the average number of words per sentence and the average number of syllables per word. A high score signifies good readability, typically characterized by short sentences and words.

\item \textbf{SERP features}: The SERP features in the dataset are binary features that denote for each entry whether a specific SERP feature is present (1) or absent (0). These features can be further divided into page SERP features and positional SERP features. Page SERP features indicate whether a certain feature is present anywhere on the page. In contrast, positional SERP features specify whether a result is included in a particular SERP feature. This can manifest in various forms. For instance, a Review could be displayed beneath a URL, or a Snippet from the URL could be featured in the Knowledge Panel. Consequently, if a positional SERP feature is present, the corresponding page SERP feature must also be present, although the converse is not necessarily true. Page SERP features also include details about the type of ads displayed on a result page. We have consolidated this information into a binary feature that describes whether ads are shown. To enhance the interpretability of the dataset, it was further enriched with the total count of SERP features for both page and positional features.

\end{itemize}

\section{Methodology}\label{methodology}

\subsection{Examined Models} \label{methodology:examined_models}
This research treats the task of predicting Click-Through Rates (CTR) as a regression problem. Based on a review of relevant literature and the characteristics of the dataset at hand, three primary categories of off-the-shelf regression models were chosen for examination: linear regression models and tree-based models, and neural network models.

Linear regression models are favored for their computational efficiency and interpretability. Ordinary least squares regression (OLS) was chosen as a baseline model, given its statistical robustness and popularity in the field. To account for potential feature interactions, an adapted version of the OLS model was utilized, incorporating interactions between two features as the product of their values (Poly2). However, to control the exponential growth in the number of features, only interactions between two features with a Pearson correlation coefficient greater than 0.05 in the training dataset were included in the model. A Ridge regression model was introduced as a third variant to address potential overfitting. The Ridge model employs L2-Norm for implicit feature selection.

Tree-based models were the second category selected, specifically, ensemble models based on decision trees. These models are advantageous as they do not require specific data distributions and perform swiftly without preprocessing while being capable of modeling feature interactions. The top three performing tree-based models for tabular data from a recent comparative benchmark—Random Forest, Gradient Boosting Decision Trees (GBDT), and XGBoost—were tested, along with CatBoost, another tree-based model that has shown strong performance in recent studies.

The third category included neural network regression models. While Deep Neural Networks (DNN) have shown outstanding performance with homogeneous data types like images, audio, and text, they have been less successful with heterogeneous, tabular data \cite{grinsztajn2022tree} in comparison to tree-based models. However, other study reports that DNNs can match or even surpass the performance of traditional machine learning models on tabular data with over 10 million samples\cite{borisov2022deep}. Given that some of the best-performing models in recent CTR prediction research incorporate neural networks, this study also tests TabNet\cite{arik2021}, a neural network structure specifically designed for tabular data, as well as Wide\&Deep\cite{cheng2016wide} and DeepFM\cite{guo2017deepfm}, which are recommended by recent CTR prediction research.

\subsection{Subset Combinations} \label{methodology:subset_modelling}
To assess the predictive power of each subset, combinations of subsets were tested using the selected models. Drawing from both the literature review in section \ref{related_works}, the position subset, highlighted in the literature review and our analysis as the most significant feature, serves as the baseline for an ablation study. When testing a model on a subset combination, all features from the combined subsets are utilized for prediction. The position subset is consistently included, with the other subsets added incrementally.

Initially, the position subset is individually combined with each of the other subsets to evaluate the additional predictive power contributed by each subset. Subsequently, the predictive power of all three subsets combined—position, keyword, and SERP features—is examined to understand their collective impact. Consequently, four subset combinations are tested on each model presented in \ref{methodology:examined_models}. To maximize the potential of each model, hyperparameter tuning was conducted where reasonable. Please find the full documentation on tested parameter ranges in appendix \ref{appendix:hyperparams_tested} and the resulting parameters in appendix \ref{appendix:hyperparams_used}.

\subsection{Evaluation Metric} \label{methodology_metric}
A common evaluation metric, the Root Mean Squared Error (RMSE), is employed for a comparative analysis of the different feature subsets across models, as defined in the equation:
\begin{equation} \label{equation:rmse}
    RMSE = \sqrt{\frac{1}{N}\Sigma_{i=1}^{N}{({y_i -\hat{y}_i})^2}}
\end{equation}
This metric effectively penalizes larger errors, which aligns with the objectives of the present business case where substantial errors can lead to significant misallocations of SEO resources. Moreover, the RMSE serves as a convenient optimization metric across all model categories, offering advantages over alternative metrics like Mean Absolute Error (MAE).

\subsection{Feature Importance Evaluation} \label{methodology:feature_importance}
To evaluate the significance of SERP features in predicting CTR and to compare them with position and keyword characteristics, this research employs model interpretation techniques. These techniques assign importance values to features in a machine learning model. A high importance value indicates a greater impact on the overall prediction compared to a feature with a lower importance value. By grouping features into subsets as described earlier, we can examine the collective importance of SERP features in relation to position and keyword features. The XGBoost model is utilized for assessing the feature importance, as it has shown superior performance.

Three techniques were employed: SHAP, permutation importance, and average gain in XGBoost splits. These were chosen for two reasons: their ability to compare the global relevance of features for the prediction and their effectiveness in providing a more holistic conclusion when used together. They also serve as a way to cross-check the results.

SHAP importances are based on Shapley explanations, a game-theoretic approach that assigns individual features a share of an aggregated outcome \cite{winter2002shapley}. It measures the average contribution of a feature to the prediction across all observations \cite{lundberg2017unified, molnar2020interpretable}. Permutation importance is calculated by randomly shuffling each feature and measuring the decrease in a model’s performance caused by the permutation \cite{breiman2001random}. Features that cause a large decrease in performance when shuffled are ones that the model heavily relies on to make predictions \cite{molnar2020interpretable}. The average gain in XGBoost’s tree splits is calculated based on the intrinsic structure of the XGBoost model by averaging the improvement of the loss function in all splits in a feature is used \cite{friedman2013elements}. According to the gain metric, features with a larger average contribution to reducing the loss function are more relevant for the overall prediction.

\section{Exploratory Data Analysis (EDA)} \label{eda}

\subsection{The Broad Impact of SERP Features} \label{eda:general_impact}
\begin{figure*}[htbp]
\centerline{\includegraphics[width=0.8\textwidth]{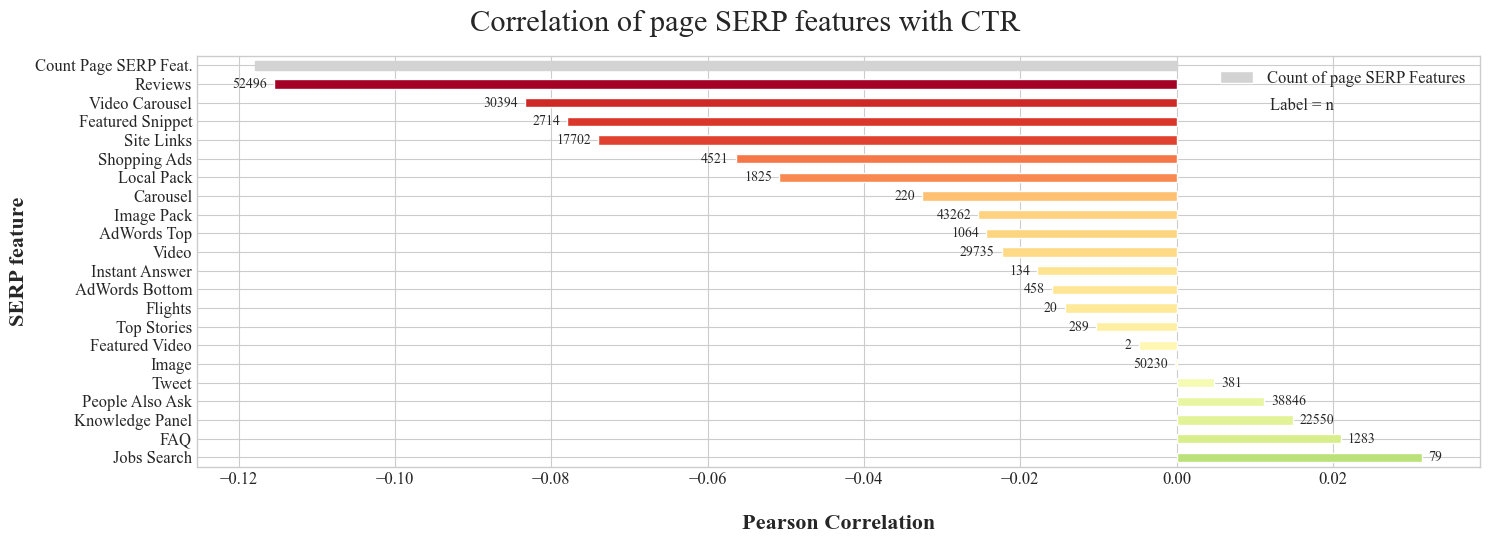}}
\caption{Correlation of SERP features with CTR when they are generally present on a result page. The color represents the strength of the effect: dark red refers to a strong negative correlation while dark green refers to a strong positive correlation.}
\label{fig:corr_serps_ctr}
\end{figure*}

The first stage of this study's exploration involves a high-level analysis of the influence of SERP features on CTR. A notable trend is observed when the count of positional SERP features, those attached to a single result, increases. This trend shows a positive correlation with CTR ($\rho=0.22$). However, an increase in the number of page SERP features appearing on the entire results page doesn't exhibit a similarly clear trend ($\rho=-0.11$), suggesting a potential negative influence on CTR. This overall negative trend, albeit slight, can be better understood by differentiating the unique SERP features, thereby illuminating the individual drivers of this influence illustrated in Figure \ref{fig:corr_serps_ctr}).

The data suggest that the presence of most SERP features correlates negatively with CTR. This implies that businesses aiming to optimize their search engine performance should potentially focus on keywords associated with fewer SERP features. However, the dataset indicates that SERP features are almost ubiquitous, with 99.8\% of the result pages showcasing at least one SERP feature and the majority displaying between four and six features. Given this prevalence, it becomes essential for website providers to comprehend the specific effects and dynamics of individual SERP features, especially those that could enhance their CTR. The impact of appearing in certain SERP features will be further examined in section \ref{eda:in_out}.

\subsection{Variations in Effects Across Positions} \label{eda:positions}
\begin{figure}[htbp]
\centerline{\includegraphics[width=\columnwidth]{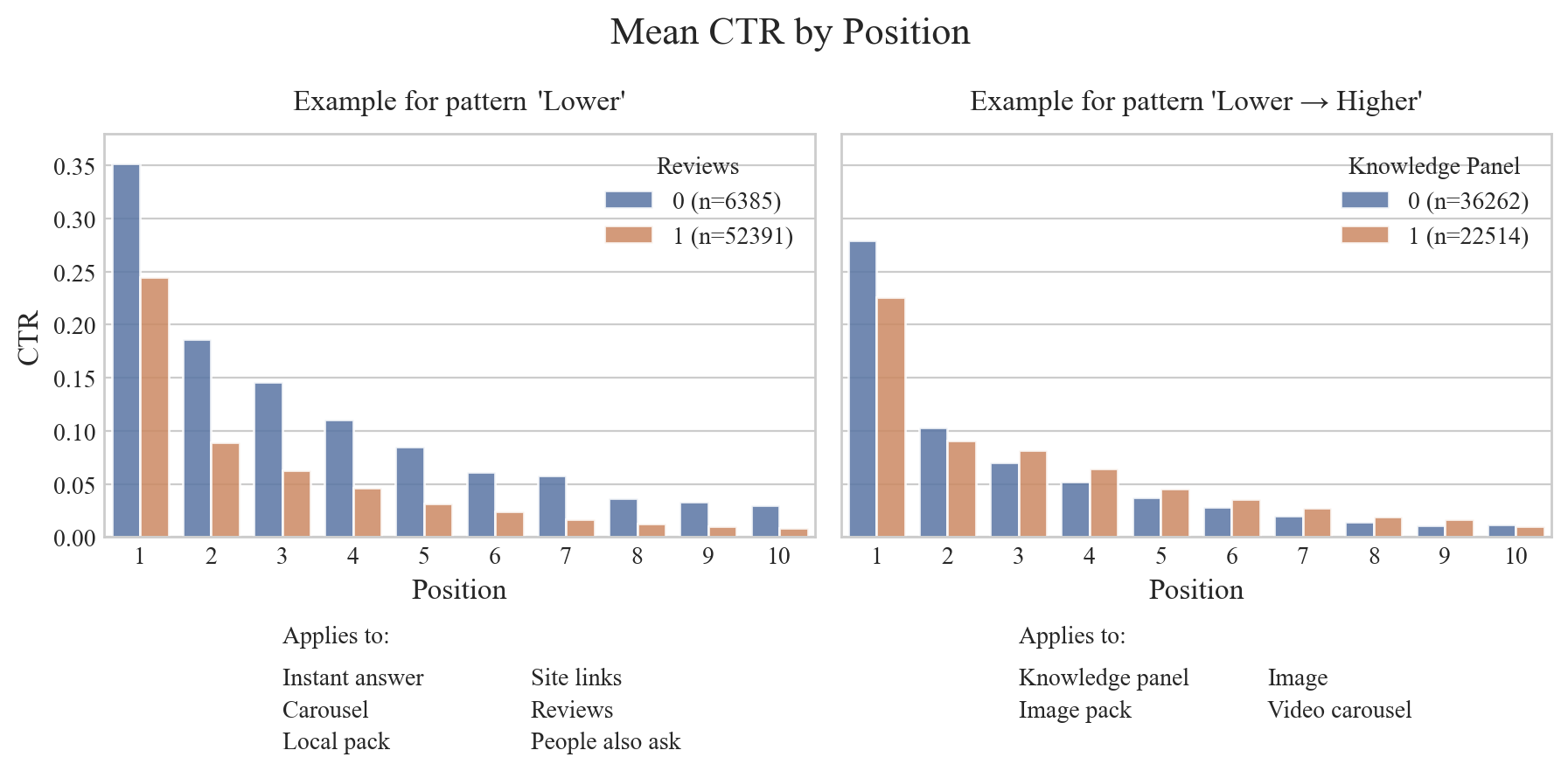}}
\caption{Mean CTR per position if a SERP feature is present (1) compared to when it is not present (0). The left plot shows the pattern ‘Lower’ and the right plot pattern ‘Lower → Higher’. SERP features without a clear pattern or $n < 1000$ are not listed.}
\label{fig:ctr_pos_lower_higher}
\end{figure}

The general analysis in section \ref{eda:general_impact} revealed a negative overall influence of the presence of SERP features on CTR. However, when incorporating the position of a result as an additional variable, intriguing trends emerge. For some SERP features, the average CTR per position remains relatively unaffected by the presence of the SERP feature. For others, the presence significantly influences the average CTR. Two primary patterns are observable: with the presence of a SERP feature, either the average CTR declines across all positions (pattern: ‘Lower’), or the average CTR decreases for the initial three positions and then increases for subsequent positions (pattern: ‘Lower → Higher’). Figure \ref{fig:ctr_pos_lower_higher} exemplifies these patterns and lists the SERP features to which they apply.

The reasons for the emergence of these patterns for specific SERP features aren't immediately evident. Nonetheless, it's worth noting that certain features have a considerable correlation with user intents. For instance, the presence of Image SERP features is strongly correlated with transactional search intent ($\rho = 0.29$). Therefore, while interpreting these results, it is crucial to bear in mind that the observed effects might be influenced by other feature categories, such as the search intents.

\begin{figure}[htbp]
\centerline{\includegraphics[width=\columnwidth]{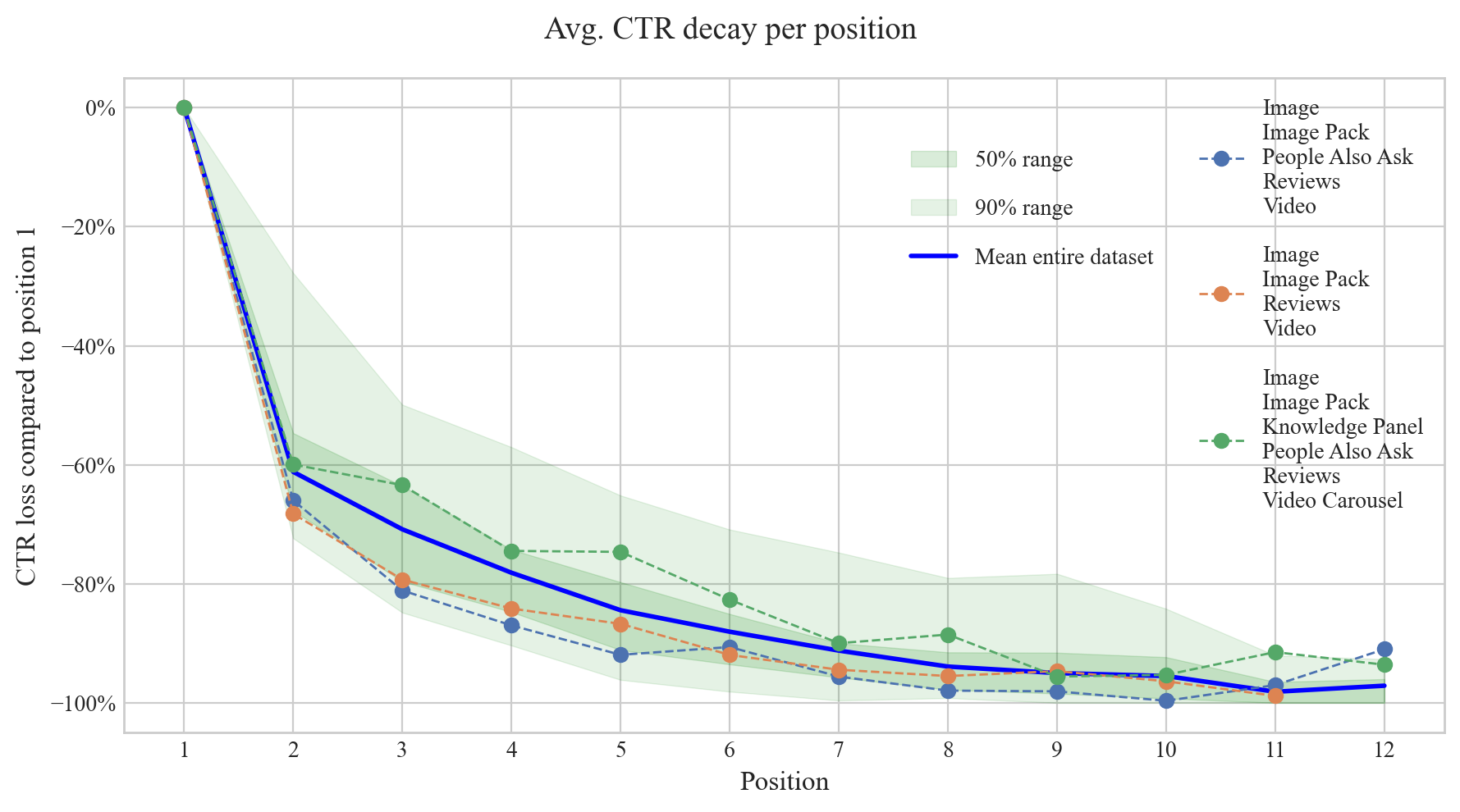}}
\caption{Loss in CTR per position. Values are the average CTR loss for each position in the percentage of average CTR on position 1. Only combinations with more than 200 occurrences are considered (n=59). Dashed lines represent the top three most frequent SERP feature combinations. Shaded areas represent the range in which 50\% and 90\% of all values fall.}
\label{fig:ctr_decay}
\end{figure}

While it is a well-established notion that a high-ranking position is vital for maximizing CTR, the detailed investigation into the impact of SERP features prompts the question of whether the importance of top-ranking positions varies in different scenarios. As discussed above, the effects of individual SERP features have been isolated for analysis. However, in real-world situations, SERP features often appear concurrently, which likely leads to inter-feature influence. To account for this, we examine the importance of positions for each combination of page SERP features. For a more comprehensive comparison, the CTR is normalized to the average value at the first position for each combination that appears more than 200 times. Subsequently, the rate of decay for each subsequent position is calculated. The analysis reveals that patterns identified for individual SERP features persist across combinations of them. Among the top three combinations by count, two show a faster decay in CTR, while the other decays much slower compared to the entire dataset. For instance, the presence of three different SERP features can double the CTR at the third position, as demonstrated by the contrast between the green and blue combinations in Figure \ref{fig:ctr_decay}.

\subsection{Differential Impact of SERP Features on Included and Excluded Results} \label{eda:in_out}
Sections \ref{eda:general_impact} and \ref{eda:positions} established that SERP features significantly affect consumer click behavior on search result pages. Initially, the investigation of SERP features was conducted at the page level, evaluating their overarching impact on all results on a page. However, it is critical to acknowledge that SERP features are not merely passive elements on a page; they actively link to some of the websites listed on the results page, thereby directly contributing to the CTR of those specific results.

To gain a comprehensive understanding of SERP features' impact, it becomes imperative to differentiate between their influence on results they link and those they do not. An illustrative example can be derived from analyzing images shown next to a result. We need to differentiate between (i) the effect the image has on the result it is linked with and (ii) the effect on a result that does not have an accompanying image, while others do. This differentiation warrants the introduction of a new set of feature categories, encapsulating whether a result is displayed within a SERP feature (termed as "Result in feature") or not included within the SERP feature ("Result not in feature"). For example, the general trend observed in Figure \ref{fig:corr_serps_ctr}, which indicates a negative correlation, encapsulates the effect of SERP features without considering whether the results are included within them or not. A more nuanced understanding can be obtained by differentiating these two possibilities.

\begin{figure*}[htbp]
\centerline{\includegraphics[width=1.9\columnwidth]{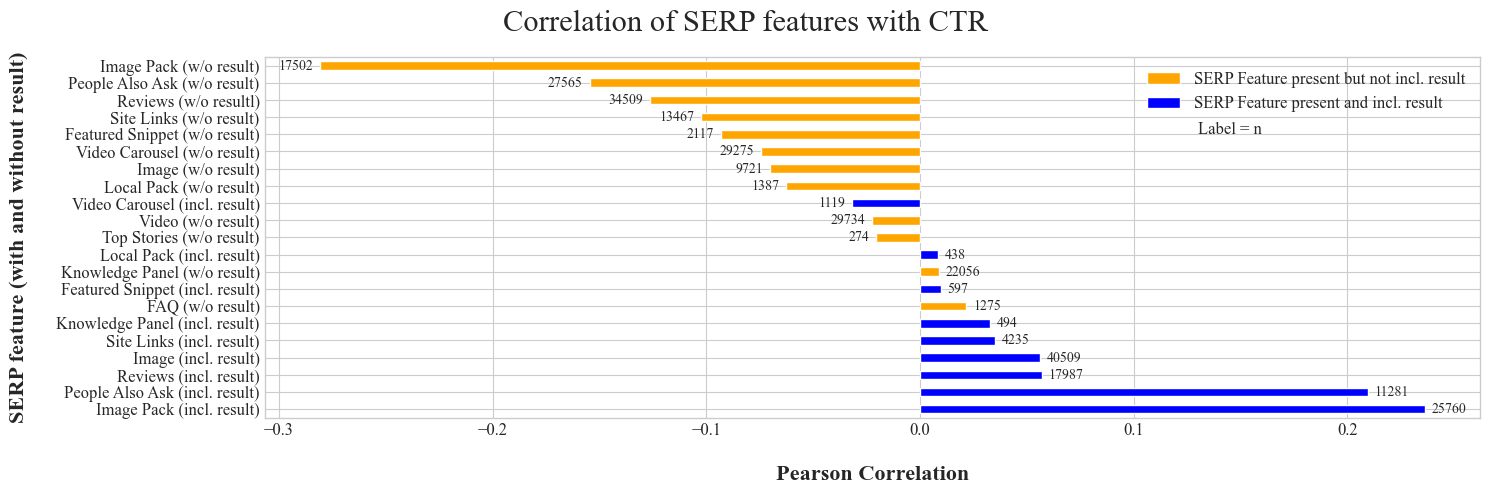}}
\caption{Correlation of SERP features with CTR differentiating between "Result in feature" and "Result not in feature". For some features, not enough results were included, which is why there are more features in the "Result not in feature" category than "Result in feature".}
\label{fig:corr_serps_ctr_in_out}
\end{figure*}

Figure \ref{fig:corr_serps_ctr_in_out} underscore the divergent effects between the two categories. Not being included in most SERP features negatively affects the CTR, while being included within a feature generally increases the CTR. This dichotomy prompts an investigation into what makes clicking on links within SERP features attractive to users, and how website providers can leverage this to their advantage. One noteworthy pattern emerges from the analysis of the Image Pack feature. This feature, representing a collection of images on the results page, exhibits a negative correlation ($\rho=-0.3$) with CTR when a result is not included. Conversely, when a result is part of the Image Pack, the correlation with CTR is positive ($\rho=0.25$). The SERP feature, "People Also Ask", exhibits a similar pattern, albeit with a less pronounced magnitude.

Further analysis reveals that the presence of SERP features tends to negatively influence the CTR of the first position, regardless of whether the result is included within the SERP feature. The positive effect of "Result in feature", as seen in Figure \ref{fig:corr_serps_ctr_in_out}, only becomes evident from the second position onward. In contrast, the effect of "Result not in feature" remains negative across all positions.

In conclusion, SERP features can have varying impacts on the CTR depending on the position of a result, regardless of whether the result is included within them. Results in the first position are particularly susceptible to decreased CTR due to increased SERP features. For results in lower positions, being featured within SERP features tends to be beneficial, while not being included in SERP features can adversely affect a result’s CTR.

\section{Evaluation and Interpretation of Model Results} \label{modeling}

\subsection{Comparative Analysis of Models} \label{modeling:model_comparison}
Table \ref{table:rmse} presents a comprehensive performance breakdown of each model and subset combination, forming the basis of our analysis in this section. In the evaluation of the model performance, a conspicuous trend surfaces. Tree-based models consistently outperform linear models and neural networks. In particular, GBDT and its variants, such as CatBoost and XGBoost, produce the most impressive results. To better understand the reduction in RMSE and the importance of the features, we decided to concentrate on a single model best suited to address the problem in question. This approach ensures that our interpretations are not skewed by outliers from models that perform poorly or are unsuitable. Because both CatBoost and XGBoost perform equally well, we use XGBoost for further analysis due to its widespread recognition and familiarity in the field.

\begin{table*}[!htb]
\centering
\begin{tabular}{lllll}
& \multicolumn{4}{c}{\textbf{RMSE of subset combinations}} \\ \cmidrule{2-5}
\textbf{Model} & \textbf{Position} & \textbf{Position + Keyword} & \textbf{Position + SERP Feat} & \textbf{Position + Keyword + SERP Feat} \\ \toprule
OLS             & 0.153             & 0.145                        & 0.145                          & 0.137                                  \\ 
Poly2           & 0.143             & 0.129                        & 0.131                          & 0.122                                  \\ 
Ridge           & 0.152             & 0.144                        & 0.144                          & 0.137                                  \\ \midrule
Random Forest   & 0.137             & 0.108                        & 0.120                          & 0.100                                  \\ 
GBDT            & \underline{0.134}             & 0.109                        & 0.122                          & 0.108                                  \\ 
XGBoost         & \textbf{0.133}             & \underline{0.104}                        & \textbf{0.120}                          & \textbf{0.098}                                  \\ 
CatBoost        & \underline{0.134}             & \textbf{0.103}                        & \textbf{0.120}                          & \underline{0.099}                                  \\ \midrule
TabNet          & 0.137             & 0.121                        & 0.130                          & 0.118                                  \\ 
Wide\&Deep      & 0.137             & 0.123                        & 0.152                          & 0.128                                  \\ 
DeepFM          & 0.136             & 0.135                        & 0.137                          & 0.133                                  \\ \bottomrule
\vspace{.1px}
\end{tabular}
\caption{Benchmark results on different feature subsets. The top results for each feature subset are \textbf{bold}. We also \underline{underline} the second-best results.}
\label{table:rmse}
\end{table*}

\subsection{Assessing the Explanatory Power of Feature Subsets} \label{modeling:subset_comparison}
By running the models on various combinations of feature subsets, we can assess the explanatory power of each subset and infer their overall relevance for CTR prediction. When evaluating the improvement in RMSE resulting from the inclusion of additional feature subsets, we focus primarily on the results from XGBoost. To further validate these comparisons, we also conduct paired t-tests on the results of the top five models on average. This allows us to determine if the observed differences in RMSE across feature subsets are statistically significant.
We observe that RMSE scores improve upon the addition of other subsets to the position subset. Notably, the inclusion of the keyword subset results in a significant reduction in RMSE ($-22.6\%$). This subset demonstrates a greater explanatory power than the SERP features subset, which further improves the RMSE ($-9.8\%$). When SERP features are added to the position + keyword combination, which was previously the best, there is an additional improvement in prediction ($-4.9\%$). All these differences are statistically significant according to the paired t-test.

Collectively, these findings suggest that all feature subsets are valuable, as they decrease error when added and enhance the baseline model that solely relies on the position feature. Consequently, we can deduce that the additional SERP features do have an impact on CTR and, by extension, user behavior. The larger reduction in RMSE for the keyword subset compared to the SERP feature subset indicates that they have a more pronounced effect on CTRs. However, this observation is merely indicative, as the importances of the features are not quantified from the perspective of the model. The next section provides a quantification of feature importances.

\subsection{Assessing the Importances of Features} \label{modeling:feature_importances}
To evaluate the relevance of the feature subsets defined in section \ref{methodology:subset_modelling}, we applied three different metrics: gain, permutation importance, and SHAP importance.
Table \ref{table:feature_subset_importance} presents the comparison of feature subset importance as per different measures, and corroborates our previous findings. The position subset consistently demonstrates the highest importance score across all metrics, followed by the keyword and SERP feature subsets. Regarding attributing feature importances, SHAP importance offers the most balanced perspective, with all subsets having an importance value between 28.9\% and 36.7\%. In contrast, permutation importance presents the most variance, with the 'position' subset receiving 55.9\% and 'SERP features' only 16.2\%.

\begin{table}[!htb]
\centering
\begin{tabular}{llll}
\textbf{}               & \multicolumn{3}{c}{\textbf{Importance Measure}}                                                                       \\ \cmidrule{2-4} 
\textbf{Feature Subset} & \textbf{Gain} & \textbf{\begin{tabular}[c]{@{}l@{}}Permutation \\ Importance\end{tabular}} & \textbf{SHAP Importance} \\ \toprule
Position                & 44.7\%        & 55.9\%                                                                     & 36.7\%                   \\ \midrule
Keyword                 & 35.9\%        & 27.9\%                                                                     & 34.4\%                   \\ \midrule
SERP Features           & 19.3\%        & 16.2\%                                                                     & 28.9\%                   \\ \bottomrule
\vspace{.1px}
\end{tabular}
\caption{Comparison of feature subset importance as per different measures}
\label{table:feature_subset_importance}
\end{table}

This difference can be explained by the methodology underlying each calculation. Both permutation importance and average gain rely heavily on RMSE as an evaluation metric, which tends to penalize large deviations disproportionately. For instance, the 'position' feature, which has a major impact and sets a range of plausible CTR values, will be greatly affected by permutations. To illustrate, a permutation of the 'position' feature from 1 to 10 would yield a significantly different range of expectable CTRs compared to permuting the presence of a single SERP feature, which would have a much smaller impact on RMSE.

Nonetheless, the interpretation of SHAP values indicates that, on average, SERP features contribute to 28.9\% of the difference between a given prediction and the average prediction, suggesting that SERP features may have a greater significance for CTR than previously assumed. These findings, which are both data and model-driven, highlight the need for further research that could delve into the behavior of individual users and explore the significance of SERP features for non-e-commerce searches.

\section{Discussion} \label{discussion}
\subsection{Theoretical Implications} \label{discussion:theoretical_implications}
This study contributes to the knowledge of search engine optimization (SEO) practices by comparing the influence exerted by ranked positions, keyword-related characteristics, and SERP features on CTR. Our findings not only confirm the widely accepted view that the ranking position of a search engine page result greatly influences its click-through rate (CTR) but also reveal that other SERP characteristics — such as the presence of specific SERP features, the nature of the searched keyword — can considerably affect CTR. 

Our analysis, which encompasses a vast array of relevant SERP characteristics, is facilitated by an extensive and novel dataset that surpasses previous studies in its scope and comprehensiveness. This study also marks the first attempt to incorporate many diverse SERP features into a single analytical framework, thereby expanding upon existing research that typically only assesses the impact of individual SERP features. Furthermore, this research provides valuable insights into the effectiveness of various off-the-shelf machine learning models for predicting aggregated CTR. Notably, we demonstrate that easily applicable tree-based models can outperform state-of-the-art CTR prediction models.

\subsection{Implications for Practitioners} \label{discussion:practical_implications}
The findings from this study hold significant implications for SEO practitioners. While securing a high ranking remains a crucial optimization goal, our analysis also emphasizes the considerable potential of other SERP characteristics, particularly SERP features, in enhancing CTR.

The implications of SERP features can be distilled into three key insights. Firstly, SERP features typically reduce the click-through rate of results, leading to an increase in zero-click searches. Consequently, website providers should anticipate that the continual introduction of higher quality and a greater quantity of SERP features may lead to increased searches where users do not leave the SERP, thereby negatively impacting providers' revenues. Secondly, the presence of certain SERP features, especially in specific combinations, can significantly affect the importance of a high ranking on a results page. Some SERP features can either divert CTR away from or concentrate it towards the top positions. Therefore, efforts to secure a high ranking should be strategically directed towards keywords that, in conjunction with the present SERP features, yield high CTR. Lastly, appearing within SERP features emerges as a new potential focus for SEO efforts. Similar to securing a high ranking, having a website linked within a SERP feature can significantly enhance CTR, irrespective of the position of the actual result.

However, optimizing for SERP feature appearances presents a unique challenge. Although website developers can increase the likelihood of their content being selected by Google's crawlers through the appropriate use of metadata descriptions, there are no guarantees. This process is largely dependent on Google's discretion, and the algorithmic process determining feature selection remains largely opaque.

Looking forward, one potential game-changer in the SERP landscape is the integration of Large Language Models (LLM) like ChatGPT. These AI models have demonstrated their ability to surpass traditional search engines in terms of user experience and problem resolution. If Google were to incorporate its own LLM as a SERP feature or in another capacity within its search engine, this could have an even more profound impact than the findings of this study suggest. Therefore, it is imperative for practitioners to stay abreast of changes in SERP features.

\subsection{Limitations} \label{discussion:limitations}
Despite our findings, some limitations inherent to the dataset warrant mention. The data only includes US-based desktop searches. As such, we cannot conclusively state whether these findings are applicable to mobile searches or searches conducted in other countries. Although the structure of Google's search result page is largely uniform globally, user behaviors may differ across countries, and even more so across devices. Furthermore, our dataset was of medium size and did not allow for the analysis of rare combinations of result page characteristics due to a limited sample size. Lastly, our findings are based on data collected between May and August 2022. However, Google's search result pages are subject to frequent changes, including the introduction and adjustment of SERP features. As a result, our findings may have limited applicability to result pages that have undergone substantial changes post-dataset collection.

\subsection{Future Research}
Future research could address the limitations of the current study by extending the scope of the dataset to include searches from other countries, mobile devices, and a larger time horizon. This would enable a more comprehensive time series analysis. In addition, future studies could delve deeper into the analysis of result page characteristics other than position and SERP features, such as keyword or result characteristics, which this study has revealed to be significant. Finally, given the importance of appearing in SERP features highlighted in our study, future work could explore the factors that determine the likelihood of a result appearing in these features.

\section{Conclusion} \label{conclusion}
This paper aims to provide an in-depth analysis of the influence that Search Engine Results Page (SERP) features exert on the Click Through Rates (CTR) of organic search engine results. The results of our study confirm that SERP features, on average, exert a negative influence on CTR. Nevertheless, it has been emphasized that the specific circumstances surrounding a result - such as its position, the specific SERP features shown, and whether a website is included in particular SERP features - can significantly modulate this influence. As such, it has been demonstrated that websites that are featured in specific SERP features, or those that are ranked lower in the results, can actually derive benefits from the presence of SERP features. This paper also provides a comparative analysis of the influence of SERP features against other result page characteristics. The findings underscore that SERP features have a tangible impact on CTR, thereby holding their own against other result page characteristics. Despite this, it remains clear that the most dominant determinant of CTR is the result's position.

Our work offers several major contributions. Primarily, we conducted a comprehensive, wide-reaching analysis using a dataset that is novel in its scope and detail. With this dataset, we have examined the impact of more than 20 different SERP features, thereby offering a holistic view of the landscape of SERP features. Additionally, we have compared the importance of SERP features to virtually all other relevant result page characteristics, thereby offering a well-rounded perspective on the factors that influence CTR.

Our findings also have strong practical applications. The insights derived from this work can be highly valuable to SEO practitioners, shedding light on the relevance of SERP features as a new dimension to consider in website optimization. The study elucidates the SERP features it is most beneficial to be featured in, and also provides insights into the specific combinations of SERP features that can either increase or decrease the importance of high ranking on a results page.

\bibliography{conference}
\bibliographystyle{IEEEtran}
\newpage
\onecolumn
\appendix

\subsection{Data Dictionary} \label{appendix:data_dict}

\begin{longtblr}[
  label = III,
  entry = none,
]{
  width = \linewidth,
  colspec = {Q[130]Q[500]Q[150]Q[130]},
  hline{2-67} = {-}{},
}
\textbf{Variable} & \textbf{Description} & \textbf{Data Source} & \textbf{Example}\\
\textit{Ads} & Any type of ad on the page. This can include ads on top of the search, at bottom, or shopping ads. Describes whether the SERP feature appears at least once on the page. & Self-engineered & 1\\
\textit{AdWords Bottom} & A series of ads (up to 4) that appear at the bottom of the first search results page. Describes whether the SERP feature appears at least once on the page. & SEO data provider & 1\\
\textit{AdWords Top} & A series of ads (up to 4) that appear at the top of the first search results page. Describes whether the SERP feature appears at least once on the page. & SEO data provider & 1\\
\textit{AMP} & AMP pages (i.e., results marked with the word "AMP" and a gray lightning bolt) shown in search results on mobile devices. & SEO data provider & 1\\
\textit{Avg. Monthly Volume} & Search volume averaged over months~ & Keywordplanner & 1312.65\\
\textit{Branded} & If the keyword includes a brand name. & Grips & 1\\
\textit{Carousel} & A row of horizontally scrollable images displayed at the top of search results.~ Describes whether the SERP feature appears at least once on the page. & SEO data provider & 1\\
\textit{Clicks} & Number of clicks on targeted search result & Grips & 25\\
\textit{Competition} & Measure of competition on search term with 0.0 being the lowest and 1.0 being the highest. & Keywordplanner & 0.22\\
\textit{Count Page SERP Feat.} & Count of positive (=1) page SERP features for search result page. & Self-engineered & 2\\
\textit{Count Positional SERP Feat.} & Count of positive (=1) positional SERP features for search result. & Self-engineered & 5\\
\textit{CPC} & Cost per click of ad entry. & Keywordplanner & 2.34\\
\textit{CTR} & Click-through rate (clicks divided by impressions). & Grips & 0.25\\
\textit{Domain} & Domain of the URL of each search result. & SEO data provider & ‘jomashop.com’\\
\textit{FAQ} & A list of questions related to a particular search that shows up for a particular organic search result. When clicked on, each of the questions reveals the answer. Describes whether the SERP feature appears at least once on the page. & SEO data provider & 1\\
\textit{FAQ (Positional)} & A list of questions related to a particular search that shows up for a particular organic search result. When clicked on, each of the questions reveals the answer.~ Describes whether the result is featured in the SERP feature. & SEO data provider & 1\\
\textit{Featured Images} & A collection of images is usually displayed at the top of the SERP if Google considers visual results to be more relevant than text results. Only for mobile devices. Describes whether the SERP feature appears at least once on the page. & SEO data provider & 1\\
\textit{Featured Snippet} & A short answer to a user's search query with a link to the third-party website it is taken from that appears at the top of all organic search results. Describes whether the SERP feature appears at least once on the page. & SEO data provider & 1\\
\textit{Featured Snippet (Positional)} & A short answer to a user's search query with a link to the third-party website it is taken from that appears at the top of all organic search results. Describes whether the result is featured in the SERP feature. & SEO data provider & 1\\
\textit{Featured Video} & A video result to a search query that is displayed at the top of all organic search results. Describes whether the SERP feature appears at least once on the page. & SEO data provider & 1\\
\textit{Flights} & A block that displays flights related to a search query. Flight results include information on flight dates, duration, the number of transfers and prices. Data is taken from Google Flights. Describes whether the SERP feature appears at least once on the page. & SEO data provider & 1\\
\textit{Hotels Pack} & A block that displays hotels related to a search query. Hotel results include information on prices and rating, and allows users to check availability for certain dates. Describes whether the SERP feature appears at least once on the page. & SEO data provider & 1\\
\textit{Image} & An image result with a thumbnail displayed along with other organic search results. Describes whether the SERP feature appears at least once on the page. & SEO data provider & 1\\
\textit{Image (Positional)} & An image result with a thumbnail displayed along with other organic search results. Describes whether the result is featured in the SERP feature. & SEO data provider & 1\\
\textit{Image Pack} & A collection of images related to a search query that is usually displayed between organic search results. Describes whether the SERP feature appears at least once on the page. & SEO data provider & 1\\
\textit{Image Pack (Positional)} & A collection of images related to a search query that is usually displayed between organic search results. Describes whether the result is featured in the SERP feature. & SEO data provider & 1\\
\textit{Impressions} & Impressions of the result, including estimates for results on the second search page.. & Grips & 100\\
\textit{Instant Answer} & A direct answer to a user's search query that is usually displayed at the top of organic search results in the form of a gray-bordered box. Describes whether the SERP feature appears at least once on the page. & SEO data provider & 1\\
\textit{Intent: Commercial} & Trying to learn more before making a purchase decision (e.g. “Subaru vs. Nissan”) & SEO data provider & 1\\
\textit{Intent: Informational} & Trying to learn more about something (e.g., “What’s a good car?”) & SEO data provider & 1\\
\textit{Intent: Navigational} & Trying to find something (e.g., “Subaru website”) & SEO data provider & 1\\
\textit{Intent: Transactional} & Trying to complete a specific action (e.g., “buy Subaru Forester”) & SEO data provider & 1\\
\textit{Jobs Search} & A number of job listings related to a search query that appear at the top of the search results page. Job listings include the job title, the company offering the job, a site where the listing was posted, etc. Describes whether SERP feature appears at least once on the page. & SEO data provider & 1\\
\textit{Keyword} & The searched keyword. & SEO data provider & ‘jomashop burberry scarf’\\
\textit{Keyword Complexity Score} & Flesch reading ease score (Flesch 1948) of the keyword, with higher values indicating easier-to-read keywords. & Self-engineered & 112\\
\textit{Keyword Count} & Number of words the search query (keyword) consists of. & Grips & 3\\
\textit{Keyword Difficulty} & Difficulty to rank for given keyword expressed from 0 to 100 & SEO data provider & 87\\
\textit{Keyword Included in Domain} & Keyword Included in Domain. & Self-engineered & 0\\
\textit{Keyword Included in URL} & Keyword Included in URL. & Self-engineered & 1\\
\textit{Keyword Length} & Number of characters in keyword & Grips & 23\\
\textit{Knowledge Panel} & Panel on right side of results~ that often includes images, facts, social media links, and other relevant information to the search query. Describes whether SERP feature appears at least once on the page. & SEO data provider & 1\\
\textit{Knowledge Panel (Positional)} & Panel on right side of results~ that often includes images, facts, social media links, and other relevant information to the search query. Describes whether the result is featured in the SERP feature. & SEO data provider & 1\\
\textit{Local Pack} & Embedded Google Maps frame on top of search results. Describes whether SERP feature appears at least once on the page. & SEO data provider & 1\\
\textit{Local Pack (Positional)} & Embedded Google Maps frame on top of search results. Describes whether the result is featured in the SERP feature. & SEO data provider & 1\\
\textit{Number of Results} & Total number of results for search of keyword & SEO data provider & 456789\\
\textit{People Also Ask} & A series of questions that may relate to a search query that appears in an expandable grid box labeled "People also ask" between search results. Describes whether SERP feature appears at least once on the page. & SEO data provider & 1\\
\textit{People Also Ask (Positional)} & A series of questions that may relate to a search query that appears in an expandable grid box labeled "People also ask" between search results. Describes whether the result is featured in the SERP feature. & SEO data provider & 1\\
\textit{Position} & Position of the result among other search results & SEO data provider & 3\\
\textit{Position (monthly)} & Average position in the last month. & Grips & 2.45\\
\textit{Position Difference} & Position subtracted from previous position, i.e. positive values mean a decrease in position & SEO data provider & -1\\
\textit{Previous Position} & Position last month & SEO data provider & 2\\
\textit{Reviews} & Organic search results marked with star ratings and including the number of reviews the star rating is based on. Describes whether SERP feature appears at least once on the page. & SEO data provider & 1\\
\textit{Reviews (Positional)} & Organic search results marked with star ratings and including the number of reviews the star rating is based on. Describes whether the result is featured in the SERP feature. & SEO data provider & 1\\
\textit{Search Volume} & The search volume of a keyword, cleaned from the original (inflated) value. & Grips & 75643\\
\textit{SERP Features by Keyword} & SERP features listed by ID, before one-hot-encoded. & SEO data provider & 1\\
\textit{Shopping Ads} & A row of horizontally scrollable paid shopping results that appear at the top of a search results page for a brand or product search query, and include the website's name, pricing, and product image. Describes whether SERP feature appears at least once on the page. & SEO data provider & 1\\
\textit{Site Links} & A set of links to other pages of a website that is displayed under the main organic search result and for brand-related search queries. Describes whether SERP feature appears at least once on the page. & SEO data provider & 1\\
\textit{Site Links (Positional)} & A set of links to other pages of a website that is displayed under the main organic search result and for brand-related search queries. Describes whether the result is featured in the SERP feature. & SEO data provider & 1\\
\textit{Top Stories} & A card-style snippet presenting up to three news-related results relevant to user's search query, which is usually displayed between organic search results. Describes whether SERP feature appears at least once on the page. & SEO data provider & 1\\
\textit{Top Stories (Positional)} & A card-style snippet presenting up to three news-related results relevant to user's search query, which is usually displayed between organic search results. Describes whether the result is featured in the SERP feature. & SEO data provider & 1\\
\textit{Trends} & How much interest web searchers have shown in a given keyword in the last 12 months. & SEO data provider & 1\\
\textit{Tweet} & A card-style snippet displaying the most recent tweets related to a search query. Describes whether SERP feature appears at least once on the page. & SEO data provider & 1\\
\textit{URL} & The url of a google search result & SEO data provider & ‘https://www.jomashop.com/ burberry-4031051.html’\\
\textit{Video} & Video results with a thumbnail displayed along with other organic search results. Describes whether SERP feature appears at least once on the page. & SEO data provider & 1\\
\textit{Video (Positional)} & Video results with a thumbnail displayed along with other organic search results.~ Describes whether the result is featured in the SERP feature. & SEO data provider & 1\\
\textit{Video Carousel} & A row of horizontally scrollable videos displayed among search results. Describes whether SERP feature appears at least once on the page. & SEO data provider & 1\\\hline
\textit{Video Carousel (Positional)} & A row of horizontally scrollable videos displayed among search results. Describes whether the result is featured in the SERP feature. & SEO data provider & 1 \\\hline
\end{longtblr}

\subsection{Feature Subsets} \label{appendix:subsets}

\begin{longtblr}[
  label = none,
  entry = none,
]{
  width = \linewidth,
  colspec = {Q[129] | Q[207]Q[301]Q[299]},
  hline{2-5} = {-}{},
  cell{1}{2} = {c=3}{c}
}
\textbf{Subset} & \textbf{Features} &  & \\
Position & Position & Position Difference & Position (monthly)\\
Keyword & {Number of Results\\Keyword Difficulty\\Intent: Commercial\\Intent: Informational\\Intent: Navigational} & {Intent: Transactional\\Competition\\CPC\\Search Volume\\Avg. Monthly Search Vol.} & {Keyword Length\\Keyword Count\\Branded\\Complexity Score}\\
SERP Feat. & {Instant Answer\\Knowledge Panel\\Carousel\\Local Pack\\Top Stories\\Image Pack\\Site Links\\Reviews\\Tweet\\Video\\Featured Video} & {Featured Snippet\\Image\\Jobs Search\\Video Carousel\\People Also Ask\\FAQ\\Flights\\Knowledge Panel (Positional)\\Local Pack (Positional)\\Top Stories (Positional)\\Image Pack (Positional)} & {Site Links (Positional)\\Reviews (Positional)\\Video (Positional)\\Featured Snippet (Positional)\\Image (Positional)\\Video Carousel (Positional)\\People Also Ask (Positional)\\FAQ (Positional)\\Ads\\Count Page SERP Feat.\\Count Positional SERP Feat.}\\
Result & Keyword in Domain & Keyword in URL & Domain\textit{ (One-Hot-Encoded)}
\end{longtblr}

\subsection{Tested Hyperparameters} \label{appendix:hyperparams_tested}

\begin{longtblr}[
  label = IV,
  entry = none,
]{
  width = \linewidth,
  colspec = {Q[131]Q[207]Q[574]Q[25]},
  hline{2,10,20,30,35,38,43} = {1-3}{},
}
\textbf{Model} & \textbf{Parameter} & \textbf{Values considered for hyperparameter tuning} & \\
Random Forest & Max depth & Categorical: [None, 2, 3, 4], p=[0.7, 0.1, 0.1, 0.1] & \\
 & Number of estimators & Integer Log Uniform: 10 → 3000 & \\
 & Criterion & Categorical: [‘squared\_error’, ‘absolute\_error’] & \\
 & Max features & Categorical: ['sqrt', 'log2', None, 0.1, 0.2, 0.3, 0.4, 0.5, 0.6, 0.7, 0.8, 0.9] & \\
 & Min samples split & Categorical: [2, 3], p=[0.95, 0.05] & \\
 & Min samples leaf & Integer Log Uniform: 1 → 50 & \\
 & Bootstrap & Categorical: [True, False] & \\
 & Min impurity decrease & Categorical: [0.0, 0.01, 0.02, 0.05], p=[0.85, 0.05, 0.05, 0.05] & \\
GBDT & Loss & Categorical: [‘squared\_error’, ‘absolute\_error’, ‘huber’] & \\
 & Learning rate & Real Log Uniform: 0.01 → 10.0 & \\
 & Subsample & Real Uniform: 0.5 → 1.0 & \\
 & Number of estimators & Integer Log Uniform: 10 → 1000 & \\
 & Criterion & Categorical: [‘friedman\_mse’, ‘squared\_error’] & \\
 & Max depth & Categorical: [None, 2, 3, 4, 5], p=[0.1, 0.1, 0.6, 0.1, 0.1] & \\
 & Min samples split & Categorical: [2, 3], p=[0.95, 0.05] & \\
 & Min samples leaf & Integer Log Uniform: 1 → 50 & \\
 & Min impurity decrease & Categorical: [0.0, 0.01, 0.02, 0.05], p=[0.85, 0.05, 0.05, 0.05] & \\
 & Max leaf nodes & Categorical: [None, 5, 10, 15], p=[0.85, 0.05, 0.05, 0.05] & \\
XGBoost & Max depth & Integer Uniform: 1 → 11 & \\
 & Number of estimators & Integer Uniform: 100 → 1000 & \\
 & Min child weight & Integer Log Uniform: 1 → 100 & \\
 & Subsample & Real Uniform: 0.5 → 1.0 & \\
 & Learning rate & Real Log Uniform: 1e-5 → 0.7 & \\
 & Col sample by level & Real Uniform: 0.5 → 1.0 & \\
 & Col sample by tree & Real Uniform: 0.5 → 1.0 & \\
 & Gamma & Real Log Uniform: 1e-8 → 7.0 & \\
 & Lambda & Real Log Uniform: 1.0 → 4.0 & \\
 & Alpha & Real Log Uniform: 1e-8 → 100.0 & \\
CatBoost & Max depth & Integer Uniform: 3 → 10 & \\
 & Learning rate & Real Log Uniform: 1e-5 → 1.0 & \\
 & Bagging temperature & Real Uniform: 0.0 → 1.0 & \\
 & L2 leaf regression & Real Log Uniform: 1.0 → 10.0 & \\
 & Leaf estimation iterations & Integer Uniform: 1 → 10 & \\
TabNet & Number decision steps & Categorical: [3, 5, 10] & \\
 & Layer size & Categorical: [8, 16, 64] & \\
 & Learning rate & Categorical: [0.01, 0.02] & \\
WideDeep & Layer 1 size & Categorical: [64, 128, 256] & \\
 & Layer 2 size & Categorical: [64, 128, 256] & \\
 & Layer 3 size & Categorical: [None, 64, 128, 256] & \\
 & Dropout ratio & Categorical: [0.0, 0.01, 0.05] & \\
 & Embedding dimension & Categorical: [4, 16, 32] & \\
DeepFM & Layer 1 size & Categorical: [64, 128, 256] & \\
 & Layer 2 size & Categorical: [64, 128, 256] & \\
 & Layer 3 size & Categorical: [None, 64, 128, 256] & \\
 & Use batch normalization & Categorical: [True, False] & \\
 & Dropout ratio & Categorical: [0.0, 0.01, 0.05] & \\
 & Embedding dimension & Categorical: [4, 16, 32] & 
\end{longtblr}

\subsection{Used Hyperparameters} \label{appendix:hyperparams_used}

\begin{longtblr}[
  label = none,
  entry = none,
]{
  width = \linewidth,
  colspec = {Q[69]Q[107]Q[65]Q[75]Q[90]Q[100]Q[81]Q[150]Q[190]},
  hline{2} = {4-9}{},
  hline{3,11,21,31,36,39,44} = {-}{},
  cell{1}{4} = {c=6}{c},
}
 &  &  & \textbf{Used for Subset} &  &  &  &  & \\
\textbf{Model} & \textbf{Parameter} & \textbf{Default} & \textbf{Position} & \textbf{Position + Keyword} & \textbf{Position + SERP Feat} & \textbf{Position + Result} & \textbf{Position + Keyword + SERP Feat} & \textbf{Position + Keyword + SERP Feat + Result}\\
Random Forest & Max depth & None & None & None & None & None & None & None\\
 & Number of estimators & 100 & 897 & 823 & 3000 & 3000 & 175 & 516\\
 & Criterion & squared \_error & squared \_error & squared \_error & squared \_error & squared \_error & squared\_error & squared\_error\\
 & Max features & 1.0 & 0.1 & sqrt & 0.3 & 0.4 & 0.3 & 0.6\\
 & Min samples split & 2 & 3 & 2 & 2 & 3 & 2 & 2\\
 & Min samples leaf & 1 & 38 & 5 & 8 & 9 & 1 & 2\\
 & Bootstrap & True & True & False & False & True & False & True\\
 & Min impurity decrease & 0.0 & 0.0 & 0.0 & 0.0 & 0.0 & 0.0 & 0.0\\
GBDT & Loss & squared \_error & squared \_error & squared \_error & squared \_error & squared \_error & squared\_error & huber\\
 & Learning rate & 0.10 & 0.067 & 0.033 & 0.030 & 0.351 & 0.126 & 0.767\\
 & Subsample & 1.0 & 0.517 & 1.0 & 0.711 & 0.833 & 0.634 & 0.819\\
 & Number of estimators & 100 & 128 & 497 & 220 & 196 & 487 & 268\\
 & Criterion & friedman \_mse & friedman \_mse & friedman \_mse & friedman \_mse & friedman \_mse & friedman\_mse & friedman\_mse\\
 & Max depth & 3 & 2 & None & None & 4 & 5 & 2\\
 & Min samples split & 2 & 3 & 2 & 2 & 3 & 3 & 2\\
 & Min samples leaf & 1 & 24 & 50 & 1 & 10 & 24 & 9\\
 & Min impurity decrease & 0.0 & 0.010 & 0.050 & 0.050 & 0.010 & 0.010 & 0.010\\
 & Max leaf nodes & None & 15 & None & 15 & 5 & 5 & 15\\
XGBoost & Max depth & 6 & 11 & 11 & 8 & 11 & 11 & 11\\
 & Number of estimators & 100 & 406 & 689 & 603 & 786 & 460 & 714\\
 & Min child weight & 1 & 100 & 1 & 1 & 1 & 1 & 1\\
 & Subsample & 1.0 & 1.0 & 0.993 & 1.0 & 1.0 & 1.0 & 0.668\\
 & Learning rate & 0.30 & 0.015 & 0.030 & 0.035 & 0.016 & 0.039 & 0.031\\
 & Col sample by level & 1.0 & 0.50 & 0.50 & 0.50 & 0.50 & 0.917 & 0.50\\
 & Col sample by tree & 1.0 & 1.0 & 1.0 & 0.50 & 0.50 & 0.708 & 1.0\\
 & Gamma & 0.0 & 1.0e-08 & 7.7e-05 & 0.009 & 1.0e-08 & 1.0e-08 & 0.001\\
 & Lambda & 1.0 & 4.0 & 4.0 & 2.859 & 1.0 & 4.0 & 1.0\\
 & Alpha & 0.0 & 1.0e-08 & 0.080 & 0.113 & 1.0e-08 & 0.014 & 1.0e-08\\
CatBoost & Max depth & 6 & 5 & 10 & 9 & 9 & 10 & 9\\
 & Learning rate & 0.030 & 0.013 & 0.041 & 0.033 & 0.026 & 0.056 & 0.054\\
 & Bagging temperature & 1.0 & 0.885 & 1.0 & 1.0 & 0.0 & 1.0 & 0.0\\
 & L2 leaf regression & 3.0 & 6.601 & 10.0 & 10.0 & 10.0 & 10.0 & 10.0\\
 & Leaf estimation iterations & \textit{Dynamic} & 10 & 10 & 1 & 6 & 10 & 10\\
TabNet & Num decision steps & 3 & 3 & 3 & 3 & 3 & 3 & 3\\
 & Layer size & 8 & 8 & 8 & 8 & 8 & 8 & 64\\
 & Learning rate & 0.02 & 0.02 & 0.02 & 0.02 & 0.02 & 0.02 & 0.025\\
WideDeep & Layer 1 size & 256 & 256 & 256 & 256 & 256 & 256 & 256\\
 & Layer 2 size & 128 & 128 & 128 & 128 & 128 & 128 & 128\\
 & Layer 3 size & 64 & 64 & 64 & 64 & 64 & 64 & 64\\
 & Dropout ratio & 0.0 & 0.0 & 0.0 & 0.0 & 0.0 & 0.0 & 0.01\\
 & Embedding dimension & 4 & 4 & 4 & 4 & 4 & 4 & 16\\
DeepFM & Layer 1 size & 256 & 256 & 256 & 256 & 256 & 256 & 256\\
 & Layer 2 size & 128 & 128 & 128 & 128 & 128 & 128 & 128\\
 & Layer 3 size & 64 & 64 & 64 & 64 & 64 & 64 & 64\\
 & Use batch normalization & False & False & False & False & False & False & False\\
 & Dropout ratio & 0.0 & 0.0 & 0.0 & 0.0 & 0.0 & 0.0 & 0.01\\
 & Embedding dimension & 4 & 4 & 4 & 4 & 4 & 4 & 16
\end{longtblr}

\section{Tables}

\end{document}